\pdfoutput=1
\documentclass[aps, prl,twocolumn, 10pt]{revtex4-2}
\usepackage[utf8]{inputenc}
\usepackage{microtype}
\usepackage{newtxtext}
\usepackage[upint]{newtxmath}
\usepackage{eucal}
\usepackage{comment}

\usepackage{graphicx}
\usepackage{booktabs}
\usepackage{enumerate}
\usepackage{siunitx}
\usepackage{braket}
\usepackage{color}
\usepackage{soul}
\usepackage{bm}

\usepackage[colorlinks, allcolors=blue]{hyperref}
\usepackage[capitalize]{cleveref}

\usepackage[svgnames]{xcolor}

\usepackage[caption=false]{subfig}

\newcommand{\vecsig}{\boldsymbol{\sigma}}

\newcommand{\ubar}[1]{\text{\b{$#1$}}}
\newcommand{\vecr}{\boldsymbol{r}}

\let\vec\mathbf

\renewcommand{\i}{{\text{i}}}

\begin{document}
\title{Quasiclassical theory of superconducting spin-splitter effects and spin-filtering via altermagnets}
\author{Hans Gløckner Giil}
\affiliation{Center for Quantum Spintronics, Department of Physics, Norwegian \\ University of Science and Technology, NO-7491 Trondheim, Norway}
\author{Bj{\o}rnulf Brekke}
\affiliation{Center for Quantum Spintronics, Department of Physics, Norwegian \\ University of Science and Technology, NO-7491 Trondheim, Norway}
\author{Jacob Linder}
\affiliation{Center for Quantum Spintronics, Department of Physics, Norwegian \\ University of Science and Technology, NO-7491 Trondheim, Norway}
\author{Arne Brataas}
\affiliation{Center for Quantum Spintronics, Department of Physics, Norwegian \\ University of Science and Technology, NO-7491 Trondheim, Norway}

\begin{abstract}
Conducting altermagnets have recently emerged as intriguing materials supporting strongly spin-polarized currents without magnetic stray fields. We demonstrate that altermagnets enable three key functionalities, merging superconductivity and spintronics. The first prediction is a controllable supercurrent-induced edge magnetization, which acts like a dissipationless spin-splitter effect. The second and third predictions are a Cooper pair spin-splitter and a filtering effect, respectively. These effects allow for spatial separation of triplet pairs with opposite spin-polarizations and spin-selective tunneling of Cooper pairs. We derive a quasiclassical theory with associated boundary conditions that describe these phenomena and explain how they can be experimentally verified. Our results open a new path for spatial control of spin signals via triplet Cooper pairs using hybrid superconductor-altermagnet devices.
\end{abstract}

\maketitle

\textit{Introduction.---} 
The interaction between magnetic and superconducting materials is a significant research topic in condensed matter physics~\cite{bergeret_rmp_05, buzdin_rmp_05, linder_nphys_15}. In addition to revealing new fundamental quantum physics, such material heterostructures are promising in cryogenic technology applications as extremely sensitive detectors of radiation and heat \cite{bergeret_rmp_18} and circuit components for qubits and dissipationless diodes \cite{nadeem_natrevphys_23}.  

A recent development in the magnetic community is the discovery of a class of antiferromagnetic materials that feature strongly spin-polarized itinerant electrons despite the absence of a net magnetization. Such materials, known as altermagnets \cite{mazin_prx_22}, are not invariant under the combined operation of parity and time-reversal and feature a spin-split band structure with a different origin than relativistic effects such as spin-orbit coupling. Because of this, the spin-splitting in altermagnets can be very large, up to the order of an electron volt. \textit{Ab initio} calculations have identified several possible material candidates that can feature altermagnetism, including metals like RuO$_2$ and Mn$_5$Si$_3$ as well as semiconductors/insulators like MnF$_2$ and La$_2$CuO$_4$ \cite{hayami_jpsj_19, ahn_prb_19, lopez-moreno_pccp_16, smejkal_sa_20, reichlova_arxiv_20, smejkal_prx_22}. Very recent experiments have, via photoemission spectroscopy, confirmed the presence of an altermagnetic state in RuO$_2$ \cite{fedchenko2023observation}, MnTe \cite{krempasky2024altermagnetic, osumi_arxiv_23} and CrSb \cite{reimers_arxiv_23}. 
Moreover, single-domain samples of order 20 $\mu$m$^2$ have recently been obtained using MnTe \cite{aminAltermagnetismImagedControlled2024}, which is large enough \cite{feofanov_nphys_10} to perform transport measurements in layered structures through a single altermagnetic domain.

As for conventional antiferromagnets, altermagnets have no magnetic stray field and exhibit ultrafast THz spin dynamics. Analogous to ferromagnets, the large anisotropic spin-splitting of electron bands renders the spin degree of freedom more accessible. For these reasons, altermagnets have quickly become promising candidates for spintronics. Their unique anisotropic spin-splitting allows for interesting transport phenomena such as spin-polarized currents \cite{shao2021spin, bose2022tilted}, giant- and tunneling magnetoresistance \cite{vsmejkal2022giant}, anomalous Hall effect \cite{feng2022anomalous}, and spin-splitter effect  \cite{gonzalez2021efficient, karube2022observation, bai2022observation, boseTiltedSpinCurrent2022}.

\begin{figure}[t!]
\includegraphics[width=0.99\columnwidth]{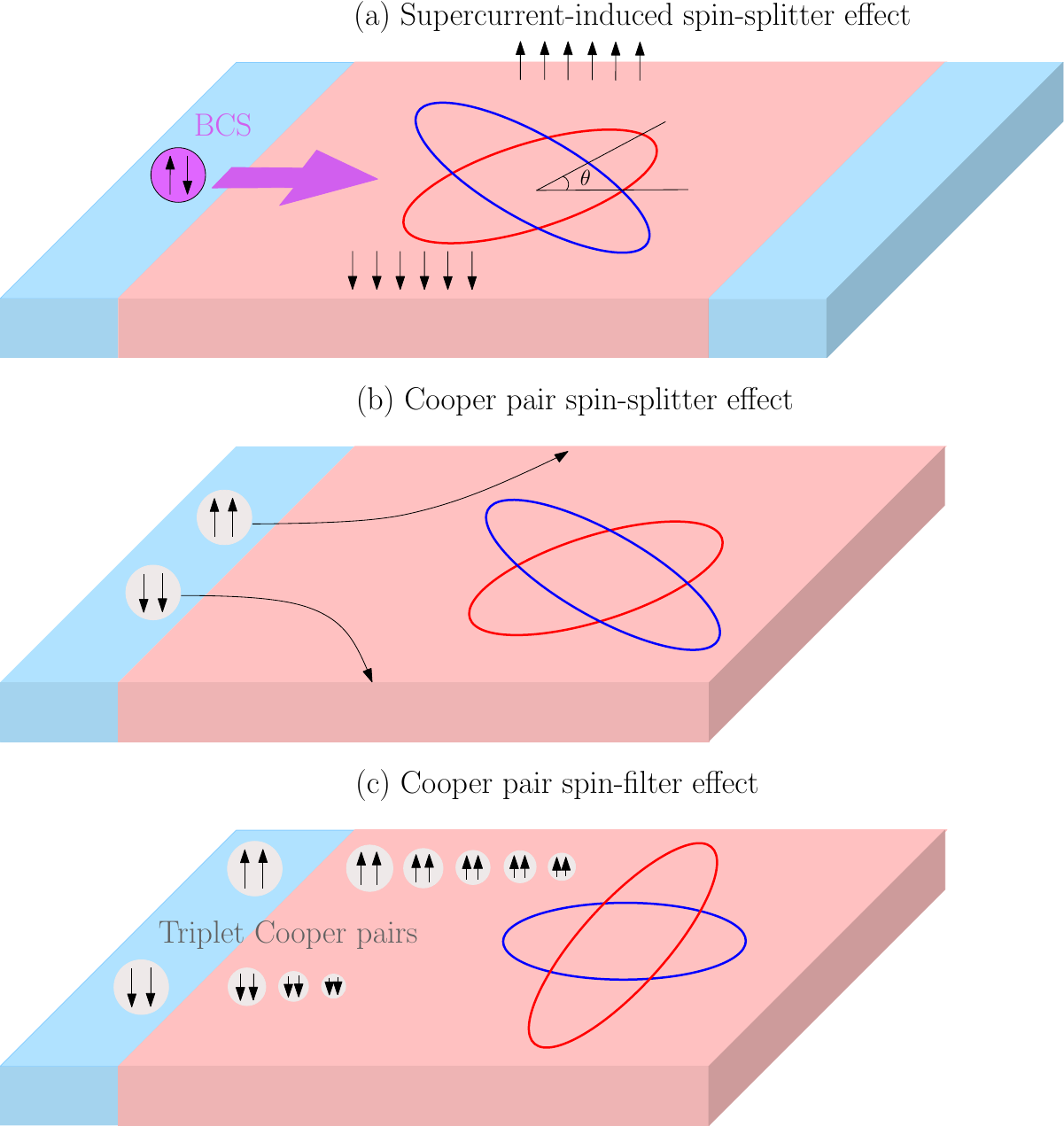}
\caption{\label{fig:model} (Color online) (a) A Josephson junction consisting of two conventional BCS superconductors separated by an altermagnet. A supercurrent induces an edge magnetization $M_z$ due to the superconducting proximity effect. The sign of the magnetization is controllable by the applied supercurrent.
(b) Spin-polarized Cooper pairs spatially separate in an altermagnet and localize at opposite edges. (c) An altermagnet filters spin-polarized Cooper pairs. The decay length discrepancy between the $\uparrow\uparrow$ and $\downarrow\downarrow$ Cooper pairs is determined by the altermagnetic strength and orientation.}
\end{figure}

On the other hand, the anisotropic spin-split bands have dramatic effects on the pairing symmetry of an accompanying superconducting state, intrinsic or in proximity contact with another material. This interplay between superconductivity and altermagnetism has very recently started to be explored, including the Josephson effect \cite{ouassou_prl_23, zhang2024finite, beenakker_prb_23}, the superconducting phase transition in the presence of altermagnetism \cite{mazin_arxiv_22, zhu_prb_23, chakraborty_arxiv_23, giil_arxiv_23, banerjee2024altermagnetic}, magnon-induced superconductivity \cite{brekke_prb_23, maeland2024many}, and Andreev reflection \cite{sun_prb_23, papaj_prb_23}. 

The use of altermagnets in the field of superconducting spintronics is, however, still underexplored.
Here, we will show that altermagnets strongly influence the transport of spin-polarized Cooper pairs. We predict that altermagnets induce two superconducting spin-splitter effects, as shown in Fig.\ \ref{fig:model}. First, a supercurrent causes an edge-magnetization in an altermagnet, with a polarization that flips when the supercurrent is reversed. Moreover, Cooper pairs polarized along or opposite to the staggered order in the altermagnet accumulate at opposite edges of the altermagnet. We also show that altermagnets filter Cooper pairs with opposite spins.
These phenomena can generate large spin supercurrents and dissipationless magnetoelectric effects without spin-orbit interactions and magnetic stray fields.
Hence, we enable new ways to control triplet superconducting correlations spatially, offering a promising direction in the interplay between magnetism and superconductivity.
In predicting these phenomena, we develop a quasiclassical theory of superconducting correlations in altermagnets and supplement this framework with boundary conditions between altermagnets and superconductors.

\textit{Theory.---} We use the quasiclassical theory of superconductivity to describe proximity-coupled altermagnets with superconductors \cite{usadel_prl_70}.
We describe the spin splitting of the altermagnet through the effective Hamiltonian
\begin{align}
\label{eq:HAM}
     H_{\textrm{AM}} &= \sum_{\sigma \sigma'} \int d\vecr \psi^\dag_\sigma(\vecr,t) \frac{\alpha}{2m} [ \bm{\ubar \sigma} \cdot \vec{l} ]_{\sigma \sigma'} T_{ij} p_i p_j \psi_{\sigma'}(\vecr, t),
\end{align}
where $\psi_\sigma(\vecr, t)$ is the electronic wave function for spin $\sigma$ at position $\vecr$ and time $t$, $\alpha$ is a dimensionless parameter specifying the strength of the altermagnetic term, and the canonical momentum is $ \vec p = - \i \nabla$. We set $\hbar = 1$. 
Furthermore, $\vec{l}$ is the Néel vector defined as the difference between the two sublattice magnetizations, and $m$ is the electron mass. 
In this work, matrices in spin space are represented by $\underline{a}$, while matrices in Nambu space, i.e. the $4\times 4$ space of spin and particle-hole degrees of freedom, are denoted by $\hat a$.
Finally, $\check a$ is a matrix in Keldysh-Nambu space, with the appropriate causality structure contingent on whether $\check a$ is a Green's function, a self-energy, or an inverse of these \cite{kamenev_field_2011}.
The real-space orientation of the altermagnet is captured by the matrix
\begin{align}
\label{eq:T}
    T = \begin{pmatrix}
        \cos{2\theta} & \sin{2\theta} \\
        \sin{2\theta} & -\cos{2\theta}
    \end{pmatrix} \, , 
\end{align}
where $\theta$ parametrizes the orientation of the altermagnet as shown in Fig.\ \ref{fig:model}.
We emphasize that the angle $\theta$ describes the orientation of the crystal lattice, not the orientation of the magnetic moments.
The altermagnetic order effectively modifies the Eilenberger equation \cite{eilenberger_zphys_68} to become
\begin{align}
\begin{split}
    \frac \i m & \vec p_F \cdot  \nabla_{\vec R} \check g 
    + [E \hat \rho_3 - \check \Sigma_\text{tot}, \check g]_-
   - \frac{\alpha}{2m} T_{ij} p_{F,i} p_{F, j}  \left[\hat m, \check g \right]_- \\&  +  \frac{\i \alpha}{2m} T_{ij}  p_{F, i} \partial_{R_j} \left[\hat m, \check g \right]_+ + \frac {\alpha}{8m}  T_{ij} \partial_{R_i} \partial_{R_j} \left[\hat m, \check g \right]_- = 0,
   \label{eq:Eilenberger}
\end{split}
\end{align}
where $\hat m = \text{diag}\left[\vecsig \cdot \vec{l}, [\vecsig \cdot \vec{l} ]^* \right]$, $\hat \rho_3 = \mathrm{diag}(1, 1, -1, -1)$, and $E$ is the energy relative to the Fermi energy. The quasiclassical Green's function $\check g$ solves the equation.
Moreover, $\vec R$ denotes the center-of-mass coordinate, and $\vec p_{F}$ is the Fermi momentum.
We use the gradient approximation in quasiclassical theory and assume the limit of a weak altermagnet with nearly constant Fermi momentum magnitude $p_F$. In this limit, the altermagnet gives rise to the three last terms in the Eilenberger Eq.\ \eqref{eq:Eilenberger}. This is our first result.

Notably, the two last terms from the altermagnetic order in Eq.\ \eqref{eq:Eilenberger} disallow the conventional normalization condition $\check g^2 = \check 1$. Higher-order terms from intrinsic spin-orbit coupling have a similar effect \cite{konschelle_prb_15}. This prohibits the standard procedure for deriving the Usadel equation, which relies on the above normalization of $\check{g}$. However, this difficulty can be circumvented in the linearized regime, which is relevant for weakly proximity-coupled hybrid structures where the anomalous quasiclassical Green's function is small. At equilibrium, it is sufficient to consider the retarded quasiclassical Green's function. We perform the linearization $\hat g^R \approx \hat \rho_3 + \hat f$. The off-diagonal part of the deviation $\hat f$ from the normal-state solution captures singlet- and triplet superconducting correlations. The absence of the normalization $\check g^2 = \check 1$ also allows for diagonal corrections in $\hat{f}$ in the quasiclassical Green's function. In the Sup. Mat. \cite{fn}, we show that these diagonal corrections do not affect the magnetization induced by the superconducting spin-splitter effects predicted here. Hence, we do not consider these corrections in further detail in the main text.

Without loss of generality, we define the Néel vector $\vec{l}$ to be along the $\hat{z}$ direction in spin space.
In the dirty limit, we average over the Fermi surface to find the Usadel equation:
\begin{align}
\label{eq:lin_Usadel_final}
    \nonumber
    D \nabla^2_{\vec R}  \ubar f =& - 2 \i E  \ubar f - \frac{\alpha D}{2}  T_{ij} \partial_{R_i} \partial_{R_j} [\ubar \tau_3,  \ubar f]_+ 
    \\
    &
    -  \frac{\i \alpha}{8m} T_{ij} \partial_{R_i} \partial_{R_j} [\ubar \tau_3,   \ubar f]_- ,
\end{align}
where $\ubar \tau_3 = \mathrm{diag}(1, -1)$ and the diffusion constant is $D = \tau v_F^2 / 2 $ in terms of the elastic scattering time $\tau$ and the Fermi velocity $v_F$. We underline that the above equation is derived without utilizing any normalization condition, which is possible due to the assumption of a weak proximity effect \cite{konschelle_prb_15}.

The prefactors of the second and third terms on the right-hand side of Eq.\ \eqref{eq:lin_Usadel_final} scale differently with the degree of disorder. 
The leading order contribution of the altermagnet is captured by the second term on the right-hand side in Eq.\ \eqref{eq:lin_Usadel_final}. 
This term gives rise to Cooper pair spin-filtering for altermagnets oriented with $\theta=0$ and a Cooper pair spin-splitter effect for a $\theta = \pi/4$ orientation, as we will show below.
The third term on the right-hand side in Eq.\ \eqref{eq:lin_Usadel_final} is smaller by a factor $mD$, which is proportional to $\lambda_F / l$, where $\lambda_F$ is the Fermi wavelength and $l$ is the mean free path. This term gives rise to singlet-triplet mixing.

We derive appropriate tunneling boundary conditions to capture the effects of proximity-coupled superconductors and altermagnets. To this end, we modify the spin-active Kupriyanov-Lukichev (KL) boundary conditions in quasiclassical theory, which are expressed as \cite{kupriyanov_zetf_88, cottet_prb_09, eschrig_njp_15, ouassou_scirep_17}
\begin{align}
\label{eq:bdc_final_L2}
    2  \check I_R
    =  G_0 [  \check g_{L}, \check g_{R}]_- + \i G_\phi [ \check g_{R}, \hat m_x]_-.
\end{align}
Here, $\check{I}_R$ is the matrix current on the right side of the junction while $\check g_{L (R)}$ is the quasiclassical Green function on the left (right) side of the boundary, and $G_0$ is the tunneling interface transparency. The spin-mixing conductance $G_\phi$ parametrizes interfacial spin polarization in the $x$-direction.
The matrix current operator is modified for altermagnets, and reads
\begin{align}
\label{eq:matrix_curr}
    {\check I}_R = G L  \check g_{R}   ( \nabla_{ R_i} + \alpha \ubar \sigma_z T_{ij} \partial_{R_j}) \check g_{R} .
\end{align}
For the interface where the AM is on the left side, the boundary condition is obtained by letting $\check{I}_R \to  -\check{I}_L$ on the left-hand side and $\check{g}_L \leftrightarrow \check{g}_R$ on the right-hand side of the boundary condition equation.
In Eq. \eqref{eq:matrix_curr}, $G$ is the bulk conductance of the altermagnet, and $L$ is the material length.
The details of the derivation and specific boundary conditions for $\theta = 0$ and $\theta = \pi/4$ are outlined in the Supp. Mat. \cite{fn}. For concreteness, we fix the values of the interface transparency to $G_0 / G = 0.4$ and $G_\phi / G = 0.1$.

We use a singlet-triplet decomposition $\underline f = (f_s + \underline{\vec d} \cdot  \underline{ \bm \sigma} ) \i \underline{\sigma}_2$, where $f_s$ is the singlet component and the vector $\vec d$ parametrizes the triplet correlations \cite{SigristUedaRMP1991}. 
The linearized Usadel equation forms two sets of coupled differential equations. In terms of the spin-polarized Cooper pair correlations $f_{\uparrow} = \i d_y - d_x$ and $f_\downarrow = \i d_y + d_x$, we find
\begin{subequations}
\label{eq:Usadel_fup_fdn}
\begin{align}
\label{triplet}
    D \nabla^2 f_\uparrow = -2 \i E f_\uparrow  +   \alpha D T_{ij} \partial_{R_i} \partial_{R_j} f_\uparrow\\
    D \nabla^2 f_\downarrow = - 2 \i E f_\downarrow  -   \alpha D T_{ij} \partial_{R_i} \partial_{R_j}  f_\downarrow .
\end{align}
\end{subequations}
Similarly, the singlet- and triplet correlations are coupled as
\begin{subequations}
\label{eq:Usadel_fs_dz}
\begin{align}
\label{singlet_final}
    D \nabla^2 f_s = - 2 \i E f_s + \i \frac{\alpha}{4m} T_{ij} \partial_{R_i} \partial_{R_j} d_z
    \\
    \label{triplet_final}
    D \nabla^2 d_z = - 2 \i E d_z + \i \frac{\alpha}{4m} T_{ij} \partial_{R_i} \partial_{R_j} f_s.
\end{align}
\end{subequations}
Interestingly, in contrast to conventional diffusive antiferromagnetic metals \cite{FyhnPRL2023Proximity, FyhnPRB2023Quasiclassical,bobkov_prb_23}, altermagnets exhibit singlet-triplet conversion also in the diffusive limit. However, from the length scale analysis, we expect the effect to be smaller than in the clean limit \cite{ouassou_prl_23, zhang2024finite}.

\begin{figure}
    \centering
    \subfloat[]{\includegraphics[height=0.12\textheight]{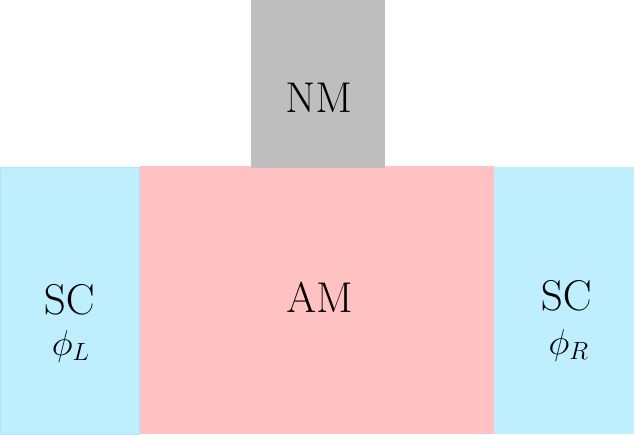}
    }
    \subfloat[]{
    \includegraphics[height=0.12\textheight]{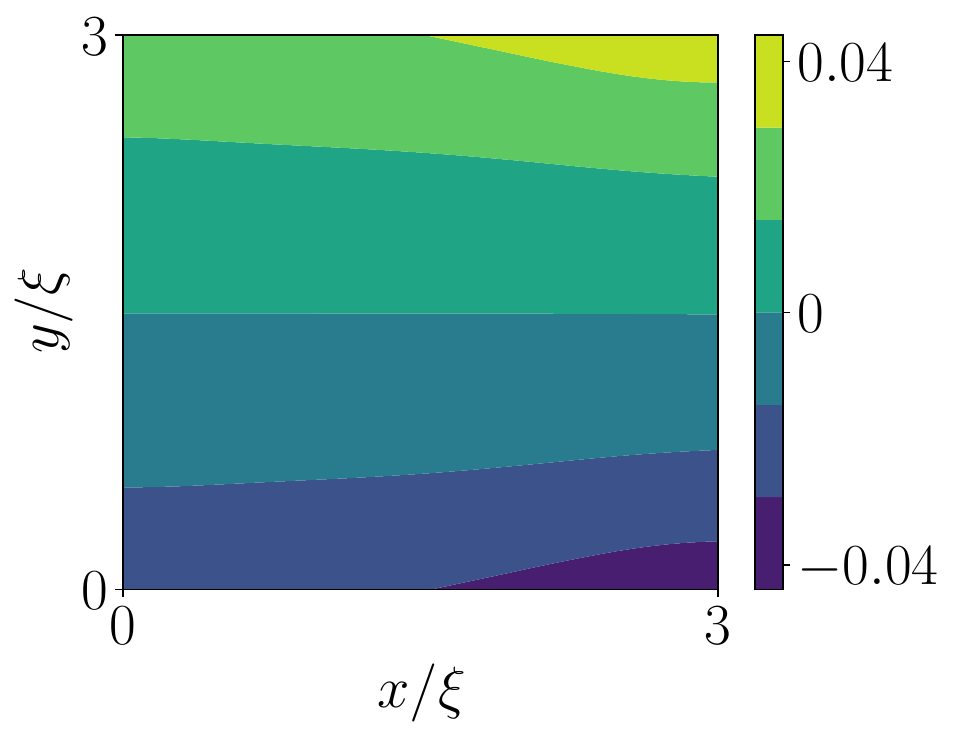}
    }
    \\
    \subfloat[]{ \includegraphics[height=0.12\textheight]{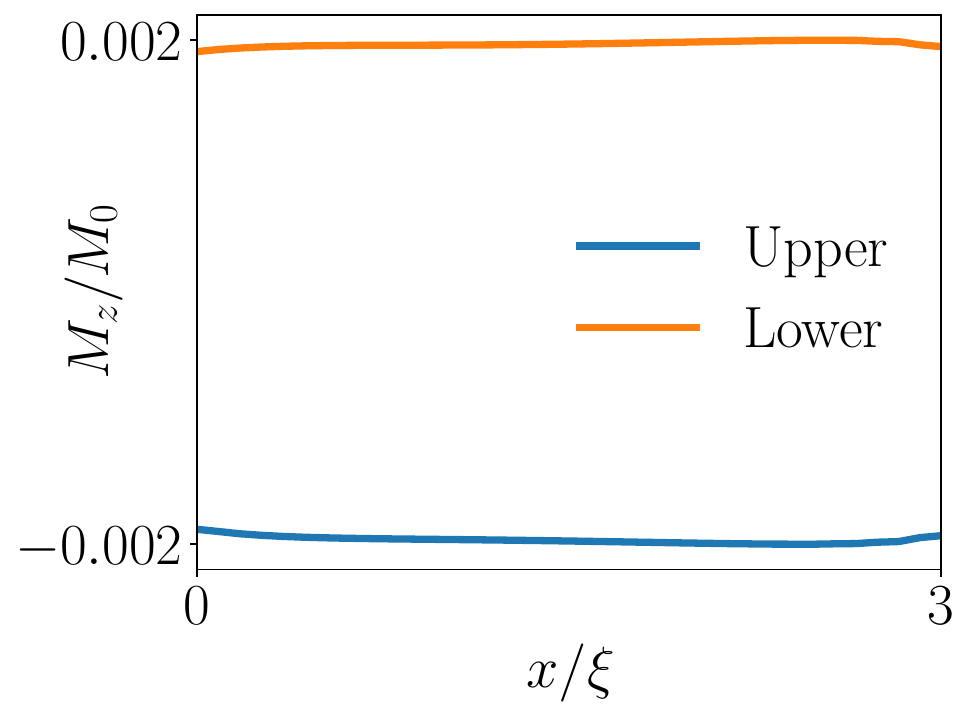}
    }
    \subfloat[]{\includegraphics[height=0.12\textheight]{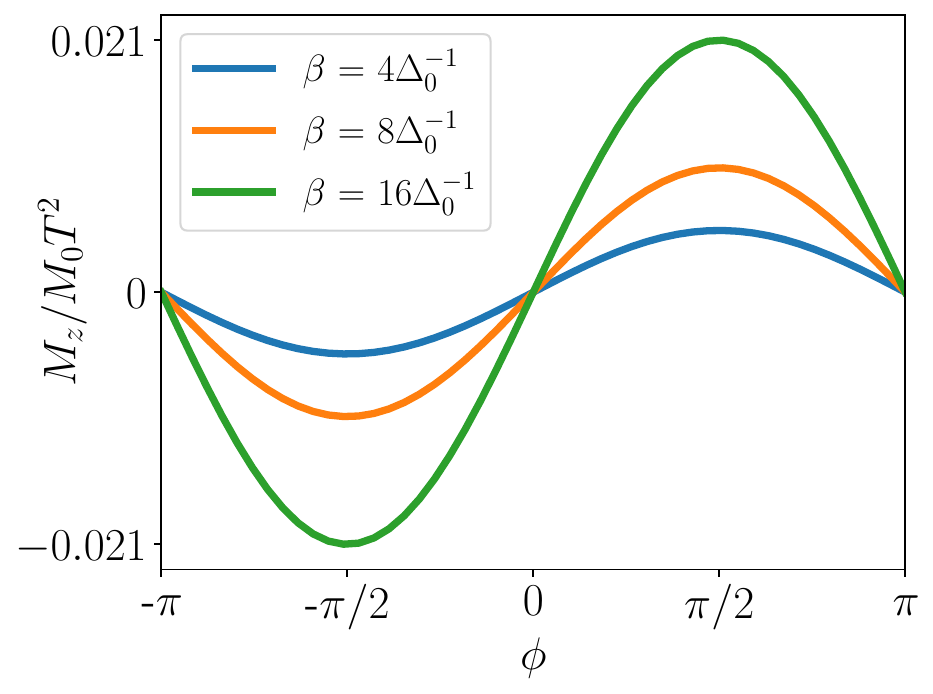}
    }
    \caption{
    (Color online) (a) Proposed experimental setup for measuring the supercurrent spin-splitter effect: attaching a normal electrode to the AM edge allows the magnetization to be measured via a spin-dependent probe such as nano-SQUID or spin-polarized STM.
    (b) The real part of the triplet correlations $d_z$ for a finite phase difference $\Delta \phi = \pi/2$ and energy $E = \Delta_0/2$.
    %Here, we use an altermagnetic strength $\alpha = 0.3$.
    % and interfacial conductance $G_0 / G = 0.4$.
    (c) The magnetization $M_z$ and the induced by the supercurrent for a phase difference $\Delta \phi = \pi/2$ in a normal metal probe connected at position $x$ along the upper and lower edges of the altermagnet in units of the diffusive coherence length $\xi = \sqrt{D/ \Delta_0}$. The inverse temperature $\beta = 16 \Delta_0^{-1}$.
    (d) The magnetization $M_z$ in the normal metal at the middle position of the upper edge in the altermagnet as a function of the phase difference $\Delta \phi$ for three different inverse temperatures.
    }
    \label{fig:Mz}
\end{figure}

We find that the altermagnet has three principal effects on the superconducting correlations. The effects depend strongly on the orientation of the altermagnet because of the $d$-wave character of the electronic spin-splitting. These effects can be experimentally observed in both the local density of states and the magnetization of the system. To compute these quantities in an unambiguous way within quasiclassical theory, we consider a normal metal coupled to the upper or lower edge of the altermagnet via a tunneling interface. Fig.\
\ref{fig:Mz}(a) shows the experimental setup. In the normal metal, $\check{g}^2=1$ holds when it is in contact with a normal reservoir (see Supp. Mat. for technical details). The proximity-induced magnetization is then computed from $M_z =  M_0\Delta_0^{-1} \int dE \text{Tr} \left[\hat \tau_3 \hat g^K \right]/ 8$. Here, $M_0 = g \mu_B N_0 \Delta_0$ and the density of states is $N=N_0\text{Re}(\text{Tr}\{\hat{g}^R\})/2$ with $N_0$ being the normal-state value. Moreover, $\Delta_0$ is the superconducting gap in the bulk superconductor.
We set the ratio between the normal metal-altermagnet interface conductivity $G_0^N$ and the bulk conductivity in the normal metal $G^N$ to be $G_0^N/ G^N L_N = 0.2 \xi^{-1}$, where $L_{\mathrm{N}}$ is the length of the normal metal and $\xi = \sqrt{D/ \Delta_0}$ is the diffusive coherence length.
We fix the altermagnetic strength $\alpha = -0.3$.

\textit{Supercurrent-induced spin-splitter effect.---}
First, we consider an altermagnet oriented with $\theta = \pi/4$ in a Josephson junction between two conventional BCS superconductors, such as Nb or Al. This setup is shown in Fig.\ \ref{fig:model}(a) and Fig.\ \ref{fig:Mz}(a).
The KL boundary conditions consistent with the altermagnetic current operator in Eq.\ \eqref{eq:matrix_curr} induce singlet-triplet mixing across the SC-AM interfaces. The resulting triplet correlations $d_z$ localize at the edges, as shown in Fig.\ \ref{fig:Mz}(b). Applying a supercurrent to the system induces a magnetization which is opposite at the upper and lower edge of the altermagnet. This can be interpreted as a dissipationless version of the spin-splitter effect \cite{gonzalez2021efficient}. We therefore call this the supercurrent spin-splitter effect.
The resulting magnetization $M_z$ in the normal metals is shown in Fig.\ \ref{fig:Mz}(c) as a function of normal metal position along the altermagnetic edge. The figure shows that the magnetization is almost constant with respect to the normal metal position. However, it has an opposite sign for the two edges, consistent with the mirror symmetry of the system.
Notably, the magnetization reverses with the supercurrent, as shown in Fig.\ \ref{fig:Mz}(d), and vanishes in the absence of a supercurrent. In total, the SC-AM-SC junction provides a controllable supercurrent spin-splitter effect evident from an accumulation of opposite magnetizations at the edges.

\textit{Cooper pair spin-splitter effect.---}
A similar effect arises for Cooper pairs that are spin-polarized along the N{\'e}el vector. We again consider an altermagnet with orientation $\theta=\pi/4$ to show this. See Fig.\ \ref{fig:model}(b). To generate triplet superconducting correlations $d_x$, we couple a conventional superconductor, like Nb or Al, to an altermagnet through a very thin ferromagnetic layer, like Co or Ni, with magnetization oriented in the $x$-direction. The magnetic interface induces a finite $d_x$ component in the altermagnet. This effect is captured by the $G_\phi$ term in Eq.\ \eqref{eq:bdc_final_L2}. 

\begin{figure}[htb]
    \centering
    \subfloat[]{\includegraphics[height=0.12\textheight]{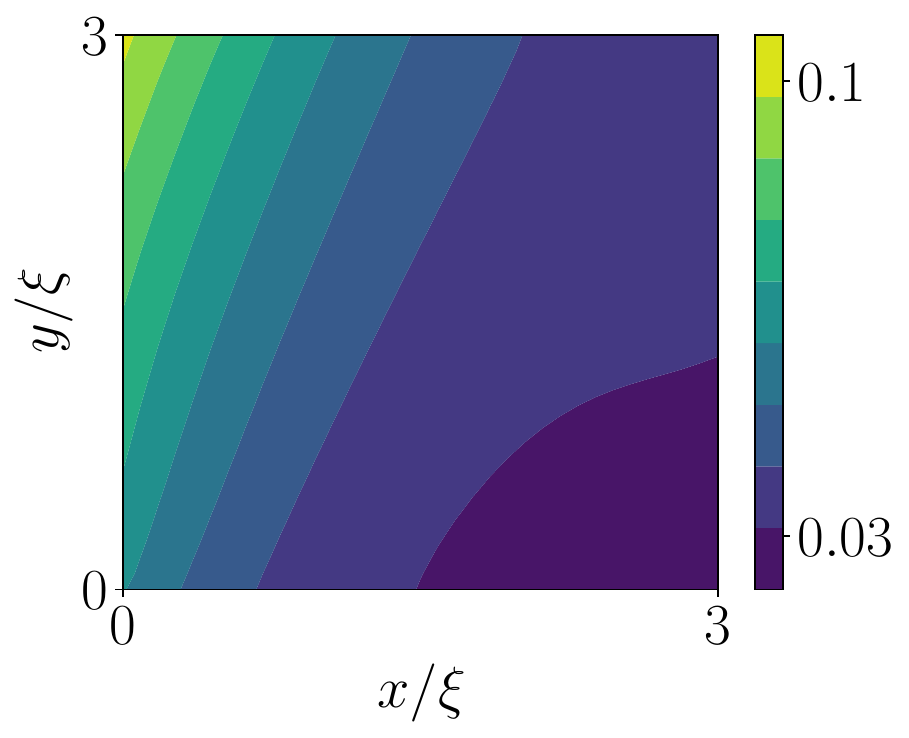}}
    \subfloat[]{\includegraphics[height=0.12\textheight]{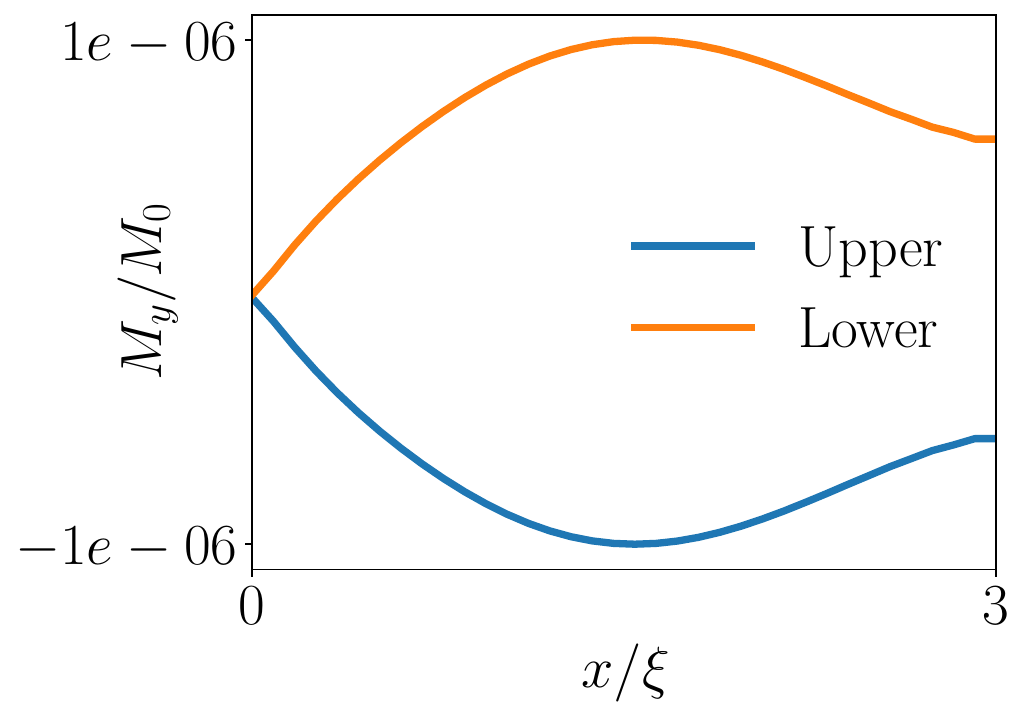}}
    \caption{(Color online) (a) The magnitude of the spin-up amplitude $f_{\uparrow}$ in the AM as a function of position, given by the coordinates $x$ and $y$. 
    %The results are given for altermagnetic strength $\alpha = 0.3$.
    (b) The magnetization $M_y$ in a normal metal that is connected at position $x$ along the upper or lower edge of the altermagnet for an inverse temperature $\beta = 16 \Delta_0^{-1}$.}
    \label{fig:cooper_pair_splitting}
\end{figure}

As shown in Eqs.\ \eqref{eq:Usadel_fup_fdn}, the Cooper pairs split and localize at opposite edges. Fig.\ \ref{fig:cooper_pair_splitting}(a) shows this effect for the spin-up amplitude. Spin-polarized Cooper pairs are generally challenging to detect experimentally \cite{yang_apl_21}. However, the Cooper pair spin-splitter effect gives rise to an associated magnetization in the $y$-direction. This can be measured in normal metals connected to the altermagnet as for the supercurrent spin-splitter effect. Fig.\ \ref{fig:cooper_pair_splitting}(b) shows the magnetization in the normal metals connected to the upper or lower edge of the altermagnet.

\begin{figure}
    \centering
    \includegraphics[width = 0.65 \linewidth]{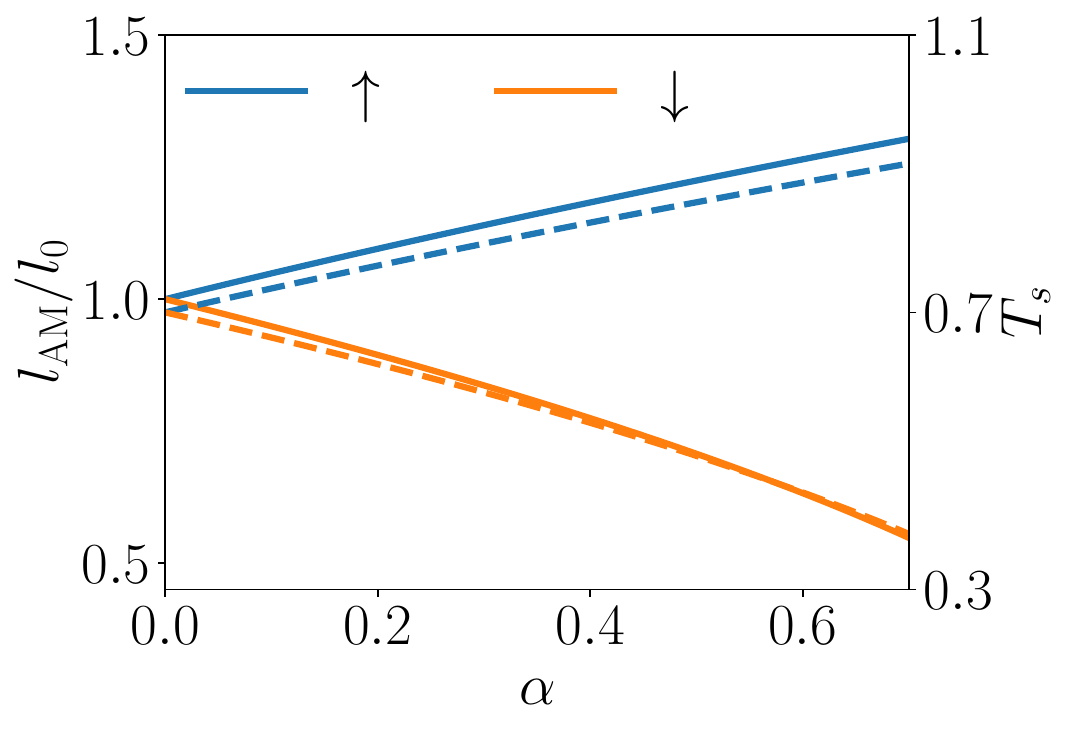}
    \caption{
    (Color online)
    The spin-polarized Cooper pair decay lengths $l_{\mathrm{AM}}$ in units of the normal metal decay length $l_0 = \sqrt{D / E}$, and the interface factor $T_s = f_s/ \sqrt{f_{\uparrow}^2 + f_\downarrow^2}$, where $s = \uparrow$($\downarrow$) for the equal spin up (down) triplet amplitudes. The results are shown as a function of the altermagnetic strength $\alpha$.
    The dashed lines correspond to the interface transmission on the right axis, and the solid lines correspond to the decay lengths on the left axis.
    } 
\label{fig:perpendicular}
\end{figure}

\textit{Cooper pair spin-filtering effect.---} 
We now consider an altermagnet oriented according to $\theta=0$ in proximity to an $s$-wave superconductor with a thin ferromagnetic interface as shown in Fig.~\ref{fig:model} (c). 
The magnetization of the ferromagnet is perpendicular to the Néel vector of the altermagnet. The primary impact of the altermagnet is spin-filtering of the $f_{\uparrow}$ and $f_{\downarrow}$-correlations. This effect has two contributions. The boundary condition in Eq.\ \eqref{eq:bdc_final_L2} yields spin-dependent tunneling that effectively spin-filters the spin-polarized Cooper pairs. 
Hence, the altermagnet gives rise to a spin-selective interfacial transmission as shown in Fig.\ \ref{fig:perpendicular}. Furthermore, the altermagnetic contribution to the Usadel equation \eqref{eq:Usadel_fup_fdn} gives rise to spin-dependent decay lengths of the spin-polarized Cooper pairs. As a consequence, the altermagnet serves as an efficient Cooper pair spin filter, as shown in Fig. \ref{fig:perpendicular}.
Finally, we note that upon the completion of this manuscript, a preprint \cite{zyuzin_arxiv_24} appeared studying a similar setup as in our Fig.\ \ref{fig:model}(a).

\textit{Concluding remarks.--}
We have developed a quasiclassical theory for proximity-coupled altermagnets and superconductors. Our theory shows that altermagnets enable the spin degree of freedom in Cooper pairs to be spatially controlled, even in the limit of strong disorder. We find that the altermagnet can control superconducting correlations in three ways.
First, we find that the altermagnet can split a supercurrent to induce a finite magnetization at the edges of the altermagnet. The direction of the magnetization is opposite at the two edges. Moreover, the supercurrent controls the magnitude and direction of the magnetization. We call this effect a supercurrent-induced spin-splitter effect.
Second, we find that triplet Cooper pairs polarized along the direction of staggered spin order are spatially separated by the altermagnet. In other words, the altermagnet serves as a mediator for creating fully spin-polarized Cooper pairs of opposite spin at opposing edges. We call this the Cooper pair spin-splitter effect.
As a third effect, we find that the altermagnet can serve as a filter for spin-polarized Cooper pairs. 
All these effects can be observed experimentally by either a local magnetization measurement or spin-polarized STM. In total, we find that altermagnets enable dissipationless magnetoelectric effects in materials void of magnetic stray fields.

\begin{acknowledgments}
	This work was supported by the Research Council of Norway through Grant No.\ 323766 and its Centres
	of Excellence funding scheme Grant No.\ 262633 ''QuSpin.'' Resources provided
	by Sigma2---the National Infrastructure for High Performance
	Computing and Data Storage in Norway, project NN9577K, are gratefully acknowledged.
\end{acknowledgments}

% \appendix

\bibliography{masterref}

\end{document}

\title{Supplementary material for \textit{Quasiclassical theory of Cooper pair spin-splitter and filtering effect via altermagnets}}
\author{Hans Gløckner Giil}
\affiliation{Center for Quantum Spintronics, Department of Physics, Norwegian \\ University of Science and Technology, NO-7491 Trondheim, Norway}
\author{Bj{\o}rnulf Brekke}
\affiliation{Center for Quantum Spintronics, Department of Physics, Norwegian \\ University of Science and Technology, NO-7491 Trondheim, Norway}
\author{Jacob Linder}
\affiliation{Center for Quantum Spintronics, Department of Physics, Norwegian \\ University of Science and Technology, NO-7491 Trondheim, Norway}
\author{Arne Brataas}
\affiliation{Center for Quantum Spintronics, Department of Physics, Norwegian \\ University of Science and Technology, NO-7491 Trondheim, Norway}
\maketitle

\section{Derivation of the Eilenberger equation in an altermagnet}

\subsection{The Hamiltonian}
We describe altermagnetism by the effective Hamiltonian,
\begin{align}
\label{eq:HAM}
     H_\text{AM} &= \sum_{\sigma \sigma'} \int d\vecr \psi^\dag_\sigma(\vecr,t) \frac{\alpha}{2m} [\vecsig \cdot \vec{l} ]_{\sigma \sigma'} T_{ij} p_i p_j \psi_{\sigma'}(\vecr, t),
 \end{align}
where $\alpha$ is the dimensionless altermagnetic strength, $\vec l$ is the Néel vector orientation, and $p_i = -\i \hbar \nabla_i $ denotes the Cartesian component $i$ of the canonical momentum.
In \eq\eqref{eq:HAM} and below, we use the Einstein summation convention for Cartesian components, i.e., $T_{ij} p_i p_j = \sum_{i, j} T_{ij} p_i p_j$.
The matrix $T$ determines the real space orientation of the altermagnet.
We parametrize this orientation by the angle $\theta$ as
\begin{align}
\label{eq:T}
    T = \begin{pmatrix}
        \cos{2\theta} & \sin{2\theta} \\
        \sin{2\theta} & -\cos{2\theta}
    \end{pmatrix}.
\end{align}
In the main text, we consider the special cases of $\theta = 0$ and $\theta = \pi / 4$ in detail. These cases correspond to the terms $p_x^2 - p_y^2$ and $2 p_x p_y$ in the Hamiltonian $H_{AM}$, respectively. 
% We denote these orientations as the \textit{perpendicular} and \textit{rotated} altermagnet, respectively.
In the following, we consider the contribution from the Hamiltonian in \eq \eqref{eq:HAM} to the quasiclassical equations of motion, following the methodology of Ref. \cite{mortenSpinChargeTransport2003}.

\subsection{Heisenberg equations of motion}
In this section, we derive equations of motion for the field operators $\psi_\sigma$ and  $\psi^\dag_\sigma$. These lead to equations of motion for the Green's functions.
The Heisenberg equation of motion for the field operator $\psi_\sigma$ for a general Hamiltonian $H$ reads
\begin{equation}
\label{eq:heisenberg}
    i \hbar \partial_t \psi_\sigma(\vecr, t) = \left[ 
    \psi_\sigma(\vecr, t), H
    \right]_- ,
\end{equation}
where we denote (anti)commutators by $[A, B]_\pm = AB  \pm BA$.
We express the commutators as anticommutators by using the identity
\begin{equation}
    [A, BC]_- = [A, B]_+ C - B [A, C]_+,
\end{equation}
and use the equal-time anti-commutation relations of the fermionic field operators,
\begin{align}
    [\psi^\dag_\sigma(\vecr, t), \psi_{\sigma'}(\vecr', t)]_+ = \delta_{\sigma \sigma'} \delta(\vecr - \vecr').
\end{align}
The Hamiltonian in \eq \eqref{eq:HAM} gives rise to the altermagnetic contribution to the equations of motion
\begin{align}
    \left[ 
    \psi_\sigma(\vecr, t), H_\text{AM}
    \right]_-
    = - \frac{\alpha \hbar^2}{2m}  \sum_{\sigma'} [\vec l \cdot \vecsig]_{\sigma \sigma'} T_{ij}\partial_i\partial_j\psi_{\sigma '}(\vecr, t).
\end{align}
The equations of motion for the adjoint operators are readily obtained by taking the Hermitian conjugate
\begin{align}
     - i \hbar \partial_t \psi_\sigma^\dag(\vecr,t) = [\psi_\sigma(\vecr,t), H]_-^\dag,  
\end{align}
and we find
\begin{align}
[\psi_\sigma, H_\text{AM}]_-^\dag = 
    - 
    \frac{\alpha \hbar^2}{2m}  \sum_{\sigma'} [\vec l \cdot \vecsig]^*_{\sigma \sigma'}  T_{ij}\partial_i\partial_j \psi^\dag_{\sigma '}(\vecr, t).
\end{align}

We proceed by introducing 4-vectors in the Nambu-spin space
\begin{equation}
    \psi^\dag =
    \begin{pmatrix}
        \psi_\uparrow^\dagger & 
        \psi_\downarrow^\dagger & 
        \psi_\uparrow & 
        \psi_\downarrow
    \end{pmatrix}.
\end{equation}
They allow us to write the equations of motion in a compact form,
\begin{align}
\label{eq:eom_psi}
    \i \hbar \hat \rho_3 \partial_t \psi(\vecr, t)
    = 
    \hat H \psi(\vec r, t) \, ,
    &&  -\i \hbar  \partial_t \psi(\vecr, t) \hat \rho_3
    = 
    \psi^\dag(\vec r, t)\hat H^\dag ,
\end{align}
where the part of $\hat H$ that depends on altermagnetism is collected in the term
\begin{align}
    \hat M(\vecr, t) &=  - \frac{\alpha \hbar^2}{2m} \hat m  T_{ij}\partial_i\partial_j.
\end{align}
Here, we introduced the matrix 
\begin{align}
      \hat m = 
      \begin{pmatrix}
        [\vecsig \cdot \vec l ] & 0 \\
        0 & [\vecsig \cdot \vec{l} ]^*
    \end{pmatrix}.
\end{align}
The other contributions to the full Hamiltonian $\hat H$ are derived in Ref.\ \onlinecite{mortenSpinChargeTransport2003}.

By introducing Green's functions in Keldysh-spin-Nambu space, we can show that the equations of motion in \eq \eqref{eq:eom_psi} imply the Green's functions of motion
\begin{align}
\label{eq:eom_greens}
    \left(\i \hbar \hat \rho_3 \partial_{t_1} - \hat H(1) \right) \check G(1, 2) = \delta(1 -2) \check 1 \,, 
    &&
    \check G(1, 2) \left(\i \hbar \hat \rho_3 \partial_{t_2} - \hat H(2) \right)^\dag = \delta(1 -2) \check 1,
\end{align}
where we use the shorthand notation $(1, 2) = (\vec r_1, t_1, \vec r_2, t_2)$, the matrix $\hat \rho_3 = \text{diag}(1, 1, -1, -1)$, and $\check 1$ is a unit matrix in Keldysh-spin-Nambu space.

The way forward is to subtract the two equations of motion in \eq \eqref{eq:eom_greens}, and Wigner transform the result.
The altermagnetic contribution from the subtraction is
\begin{equation}
\label{eq:am_in_48}
    -\hat M(1) \check G(1,2) + \check G(1,2) \hat M(2) ,
\end{equation}
where the derivatives in $\hat M(2)$ work to the left.
We define the relative $(\vec r, t)$ and center of mass coordinates $(\vec R, T)$ as
\begin{align}
    \vec R &= \frac12 (\vec r_1 + \vec r_2) \, , && T = \frac12 (t_1 + t_2) \, , \\
    \vec r &=  (\vec r_1 - \vec r_2) \, , && t = (t_1 - t_2),
\end{align}
and define the Wigner (mixed) representation as the Fourier transform with respect to the relative coordinates,
\begin{equation}
    f(\vec R, T,  \vec p, E) = \int d \vec r dt e^{-i ( \vec p \cdot \vec r - Et)/ \hbar} f(1, 2).
\end{equation}

Fourier transforming \eq \eqref{eq:am_in_48} in the relative coordinate $\vec r$, we have to calculate
\begin{equation}
    \label{eq:am_term_gorkov}
     \frac{\alpha \hbar^2}{2m} \int d\vecr e^{-i \vec p \cdot \vec r / \hbar}  \left( 
    \hat m  T_{ij}\partial_{i, 1} \partial_{j, 1}  \check G(1,2) -  T_{ij}\partial_{i, 2} \partial_{j, 2}  \check G(1,2) \hat m 
    \right) \, ,
\end{equation} 
where we emphasize that the notation $\partial_{i,l}$ denotes the partial derivative with respect to the cartesian component $i$ of the coordinate $\vec r_l$.
When evaluating \eq \eqref{eq:am_term_gorkov}, it is useful to transform the derivatives into the center-of-mass and relative coordinate,
\begin{align}
    \partial_{i, 1} \partial_{j, 1} &= \left( 
    \partial_{r_i} \partial_{r_j} + \frac12 \left[ \partial_{R_i} \partial_{r_j}+ \partial_{r_i} \partial_{R_j} \right] 
    + \frac14 \partial_{R_i} \partial_{R_j}
    \right),
    \\
     \partial_{i, 2} \partial_{j, 2} &= \left( 
    \partial_{r_i} \partial_{r_j} 
    - \frac12 \left[ \partial_{R_i} \partial_{r_j}+ \partial_{r_i} \partial_{R_j} \right] 
    + \frac14 \partial_{R_i} \partial_{R_j}
    \right)\, .
\end{align}
Inserting this into \eq \eqref{eq:am_term_gorkov}, we find three terms :
\begin{align}    \label{eq:am_term_gorkov_wigner}
    -\frac{\alpha}{2m}  T_{ij} p_{i} p_{j} \left[\hat m, \check G(\vec R, \vec p, E) \right]_- 
    +\frac{\i \alpha \hbar}{2m}   T_{ij} p_{i} \partial_{R_j}  \left[\hat m, \check G(\vec R, \vec p, E) \right]_+ 
    + \frac{\alpha \hbar^2}{8m} T_{ij} \partial_{R_i} \partial_{R_j}
    \left[\hat m, \check G(\vec R, \vec p, E) \right]_- ,
\end{align}
where we used partial integration to effectively replace $\partial_{x_i} \rightarrow i p_i/ \hbar$, pulled center-of-mass derivatives $\partial_{R_i}$ outside the integrals, and used the definition of the Wigner representation. 
Furthermore, we used the fact that $T_{ij} = T_{ji}$ to simplify the second term in \eq \eqref{eq:am_term_gorkov_wigner}.

Thus, we have found a (Gorkov) equation of motion for Green's functions (disregarding the arguments for notional convenience):
\begin{multline}
\label{eq:gorkov_final}
    \frac{\i \hbar  }{m} \vec p \cdot \nabla_{\vec R} \check G 
    + [E \hat \rho_3 - \check \Sigma_{\text{tot}} \starcom \check G]_- 
    - \frac{\alpha}{2m} T_{ij} p_{i} p_{j}  \left[\hat m, \check G \right]_- 
    +  \frac{\i  \alpha \hbar}{2m}  T_{ij} p_{i}  \partial_{R_j} \left[\hat m, \check G\right]_+
    + \frac{\alpha \hbar^2}{8m}  T_{ij} \hbar^2 \partial_{R_i} \partial_{R_j} \left[\hat m, \check G \right]_-  = 0 \, ,
\end{multline}
where the energy term comes from the time derivatives in \eq \eqref{eq:eom_greens}, and the total self-energy
$\check \Sigma_{\text{tot}}$ contains all other self-energies in the system than altermagnetism. Here, we have introduced the Moyal product $A \otimes B$, which is discussed in detail in the next section.
The factors of $\hbar$ give all quantities the dimension of energy. We set $\hbar = 1$ from here.

\subsection{Length scale analysis}
In this subsection, we discuss the length scales of our system and argue why some terms in \eq \eqref{eq:gorkov_final} can be neglected within the quasiclassical approximation. 

Let $L$ denote the characteristic length scale for spatial variations, that is, $\nabla_{\vec R} \check G \sim  1/L$.
In the quasiclassical approximation, the relevant momenta are close to the Fermi momentum, i.e. $|\vec p| \sim 1/\lambda_F$, and the length scales $\lambda_F$ associated with the momentum are much smaller than all other length scales. 
Considering the three altermagnetic terms from \eq \eqref{eq:gorkov_final}, we see that the first term is of order $1/\lambda_F^2$, the second of order $1/\lambda_F L$, and the third term of order $1/L^2$, which means that the third term should be much smaller thatn the other terms, however, we include also this term in the following to see its effects.

\subsection{Eilenberger equation}
We now apply the gradient approximation outlined above and introduce the quasiclassical Green's function into \eq \eqref{eq:gorkov_final}.
We consider the stationary case, where the circle product reduces to matrix multiplication, i.e., $A \circ B = AB$.
We truncate the gradients as explained in Ref. \cite{mortenSpinChargeTransport2003} and introduce the quasiclassical Green's function similarly as usual by defining the quantity
\begin{align}
    \xi_{\vec p } &= \frac{1}{2m} \vec p^2 - \mu .
\end{align}

The quasiclassical Green's function is sharply peaked at the Fermi surface and is formally defined as
\begin{equation}
    \check g(\vec R, T, \vec p_F, E ) = \frac \i \pi \int d \xi_{\vec p } \check G(\vec R, T, \vec p, E ). 
\end{equation}
We proceed by multiplying the Gorkov equation in \eq \eqref{eq:gorkov_final} with $\i /\pi$ and integrate over $\xi_p$ term by term:
\begin{align}
   \frac{\i}{\pi} \int d \xi_{\vec p}  \frac{i }{m} \vec p \cdot \nabla_{\vec R} \check G &= \frac \i m \vec p_F \cdot \nabla_{\vec R} \check g,
    \\
    -\frac{\i}{\pi} \frac{\alpha}{2m} \int d \xi_{\vec p}   T_{ij} p_{i} p_{j}  \left[\hat m, \check G \right]_- &= -\frac{\alpha}{2m}T_{ij} p_{F,i} p_{F, j}  \left[\hat m, \check g \right]_-,
    \\
    \frac{\i}{\pi} \frac{\i \alpha}{2m} \int d \xi_{\vec p}   T_{ij} 
    p_{i} \partial_{R_j} 
    \left[\hat m, \check G \right]_+ &=   \frac{\i \alpha}{2m}  T_{ij}  p_{F, i} \partial_{R_j}   \left[\hat m, \check g \right]_+
    \\
    \frac{\i}{\pi} \frac{\alpha}{8m}  T_{ij} \partial_{R_i} \partial_{R_j} \int d \xi_{\vec p} \left[\hat m, \check G \right]_-  
    &= 
     \frac{\alpha}{8m}  T_{ij} \partial_{R_i} \partial_{R_j} \left[\hat m, \check g \right]_-.
\end{align}
Thus, we have found the Eilenberger equation, including three altermagnetic terms:
\begin{align}
    \frac{\i}{m} \vec p_{F} \cdot \vec \nabla_{\vec R} \check g
    + [E \hat \rho_3 - \check \Sigma_{\text{tot}} , \check g]_-
    -
    \frac{\alpha}{2m} T_{ij} p_{F,i} p_{F, j}  \left[\hat m, \check g \right]_-
    +
    \frac{\i \alpha}{2m}  T_{ij} p_{F, i} \partial_{j}   \left[\hat m, \check g \right]_+
    +
    \frac{\alpha}{8m}  T_{ij} \partial_{i} \partial_{j} \left[\hat m, \check g \right]_-
    = 0.
    \label{eq:eilenberger}
\end{align}
The last two terms in \eq \eqref{eq:eilenberger} are incompatible with the usual normalization condition $\check g^2 = \check 1$
\cite{konschelle_prb_15}.

\section{Derivation of the linearized Usadel equation}
\label{section:Usadel_linearization}
In the following, we proceed by linearizing the Eilenberger equation \eqref{eq:eilenberger}  before taking the dirty limit, averaging the Eilenberger equation over the Fermi surface.
This allows us to derive a linearized Usadel equation without a normalization condition like that in Ref.\ \cite{konschelle_prb_15}.

\subsection{Linearization of the Eilenberger equation}
We consider the retarded component of \eq \eqref{eq:eilenberger},  and insert the linearization condition $\hat g^R = \hat \rho_3 + \hat f$, where $\hat f$ by assumption is a small quantity, which in general includes both diagonal and off-diagonal parts. 

Due to the general symmetries of the quasiclassical Green's function, $\hat f$ reads
\begin{align}
\label{eq:f_hat}
    \hat f = \begin{pmatrix}
        \ubar f_{d} & \ubar f_{o}\\[4pt]
        -  \tilde{\ubar f_o} & -\tilde{\ubar f_d}
    \end{pmatrix},
\end{align}
where the tilde operation is a composite energy inversion and complex conjugation, i.e., $\tilde A(E) = A^*(-E)$.
Moreover, we keep only the self-energy $\check \Sigma_\text{tot}$, and consider the upper right and the upper left $2\times 2$ parts of Eq.\ \eqref{eq:eilenberger} separately.
Crucially, the energy term only contributes to the off-diagonal part of the equations since the diagonal part drops out of the commutator
\begin{align}
[\hat \rho_3, \hat f]_- = 
    \begin{pmatrix}
        0 & 2 \ubar f_o\\
        2 \ubar{\tilde f_o} & 0
    \end{pmatrix}.
\end{align} 
We find two coupled differential equations for the $2\times 2$ matrices $\ubar f_o$ and $\ubar f_d$:
\begin{subequations}
    \begin{align}
    \nonumber
    \frac{\i}{m} \vec p_{F} \cdot \vec \nabla_{\vec R} \ubar f_o
    + 2 E \ubar f_o
    -  [\hat \Sigma_{\text{imp}}^R , \hat \rho_3 + \hat f]_-^\text{u.r.}
    -
    \frac{\alpha}{2m} T_{ij} p_{F,i} p_{F, j}  \left[\ubar \tau_3, \ubar f_o \right]_-
    &+
    \frac{\i \alpha}{2m}  T_{ij} p_{F, i} \partial_{R_j} 
    \left[\ubar \tau_3, \ubar f_o \right]_+
    \\ 
    &+
    \frac{\alpha}{8m}  T_{ij} \partial_{R_i} \partial_{R_j} \left[\ubar \tau_3, \ubar f_o \right]_- = 0 
    \label{eq:eilenberger_lin_ret_fo}
    \\ \nonumber
    \frac{\i}{m} \vec p_{F} \cdot \vec \nabla_{\vec R} \ubar f_d
    -  [\hat \Sigma_{\text{imp}}^R , \hat \rho_3 + \hat f]_-^\text{u.l.}
    -
    \frac{\alpha}{2m} T_{ij} p_{F,i} p_{F, j}  \left[\ubar \tau_3, \ubar f_d \right]_-
    &+
    \frac{\i \alpha}{2m}  T_{ij} p_{F, i} \partial_{R_j} 
    \left[\ubar \tau_3, \ubar f_d \right]_+
    \\
    &+
    \frac{\alpha}{8m}  T_{ij} \partial_{R_i} \partial_{R_j} \left[\ubar \tau_3, \ubar f_d \right]_- =0,
    \label{eq:eilenberger_lin_ret_fd}
\end{align}
\end{subequations}
where the ''u.r(l).'' superscript implies the upper right (left) $2\times 2$ part of the matrix.
In the next section, we argue that we can extract observables from only the off-diagonal equation \eqref{eq:eilenberger_lin_ret_fo}.

\subsection{The diagonal contribution}
\label{section:norm_metal}

This section proposes a setup that renders the diagonal corrections $f_d$ negligible.
The diagonal component $f_d$ arises from the lack of a normalization condition for the quasiclassical Green's function in the altermagnet. 
However, suppose we consider normal metals connected to the altermagnet, as shown in the main text in Fig. 2(a). In that case, we can derive observable quantities from the normalization condition in the normal metal. We also find that $f_d$ can be disregarded in these expressions.

In the following, we compute Green's function in the normal metal due to its proximity to the altermagnet, which hosts superconducting correlations originating from the superconductor. Thus, it has a normal and anomalous correction to its Green's function matrix. 

The anomalous Green's function in the altermagnet is known from solving Eq.\ \eqref{eq:eilenberger_lin_ret_fo}. However, the diagonal part (normal part) of the Green's function is unknown. We write the diffusive, retarded Green's function in the altermagnet as
\begin{align}
\hat{g}_A = \begin{pmatrix}
   \ubar 1+ \ubar D & \ubar A \\
    -\tilde{\ubar  A} & -\ubar 1- \tilde{\ubar  D} \\
\end{pmatrix},
\end{align}
which is a general expression of the symmetries of the quasiclassical Green's function.
We emphasize that $(\hat{g}_A)^2$ and  $\ubar D$ are unknown in the altermagnet,  but $\ubar A$ is known. 

Next, we consider the normal metal Green's function.
For a weak proximity effect (due to the tunnel coupling to the AM), we can write it as
\begin{align}
    \hat{g}_N = \hat \rho_3 + \hat{f},
\end{align}
where $\hat{f}$ must be off-diagonal due to the fact that $(\hat{g}_N)^2=1$ in the normal metal. 
As a consequence, we have
\begin{align}
    \hat{f} = \begin{pmatrix}
        0 & \ubar f \\
        -\tilde{\ubar f} & 0 \\
    \end{pmatrix}.
\end{align}
We will now show that we can determine $\hat{f}$ in the normal metal from the quantity $\ubar A$ in the altermagnet alone, allowing us to compute the density of states and magnetization in the normal metal. To do this, we consider the tunnel boundary condition between N and AM, which is given by the KL boundary condition:
\begin{align}
    2 G_N L_N \hat{g}_N \partial _x \hat{g}_N = G_0 [\hat{g}_A, \hat{g}_N],
\end{align}
where $G_N$ is the normal metal bulk conductance and $G_0$ is the interfacial conductance. 
We proceed by inserting $\hat{g}_A$ and $\hat{g}_N$ above. 
We get for the upper right block of this matrix equation that
\begin{align}
\label{bdc_nm_2}
    2 G_N L_N \partial_x\ubar f = G_0 (-2 \ubar A + 2 \ubar f + \ubar D \ubar f + \tilde{\ubar D} \ubar f).
\end{align}
The terms $\ubar A$, $\ubar D$, and $\ubar f$ are small. They are corrections in the linearization procedure. In effect, $\ubar D\ubar f$ should be negligible compared to $\ubar A$ and $\ubar f$ alone because it is second order in small quantities. Here, $\ubar D$ is the correction to $\hat \rho_3$ for the diagonal part of the Green's function in the altermagnet. It should be small in the limit of weak altermagnetism, and $f$ is the correction to the anomalous Green's function in the normal metal, which is also small since it is tunnel coupled to the AM. In contrast, $A$ is a first-order correction to the off-diagonal part of the Green's function in the altermagnet due to its proximity to the superconductor. Thus, we approximate Eq. \eqref{bdc_nm_2} as
\begin{align}
\label{bdc_nm_3}
    G_N L_N\partial_x \ubar f = - G_0 (\ubar A + \ubar f).
\end{align}
Consequently, the anomalous Green's function in an N tunnel coupled to the AM is independent of the diagonal corrections $\ubar D$ to the Green's function in the AM.

\subsubsection{Derivation of the amplitude in the normal metals}
In the limit of a semi-infinite normal metal, the solution for the anomalous Green's function is 
\begin{align}
\label{nm_sol}
    \ubar f_N = \ubar C e^{- k y}.
\end{align}
More specifically, we assume that the normal metal is longer than the diffusive coherence length and that the interface normal is parallel to the $ y$ direction.
Here, $k = \sqrt{- 2 \i E / D}$ describes the decay rate of the solution.
The KL boundary condition in the normal metal is given in Eq. \eqref{bdc_nm_3}, and becomes
\begin{align}
     G_N L_N \partial_y \ubar f_N =  - G_0 ( \ubar f_N + \ubar f_{AM}).
\end{align}
We set the factor $G_0 / (G_N L_N) \equiv \Omega = 0.2 \xi^{-1}$, and insert the solution from \eq \eqref{nm_sol} to find
\begin{align}
    - k C e^{- k y} = - \Omega( \ubar C e^{-k y} + \ubar f_{AM}), 
\end{align}
which we can solve for $C$ at $y=0$:
\begin{align}
\label{eq:C}
    \ubar C = \frac{ \Omega \ubar f_{AM}}{k - \Omega}.
\end{align}

In the next section, we take the diffusive limit and derive the Usadel equation from the linearized Eilenberger. Since we can disregard diagonal corrections in our calculation of observable quantities, we drop the ''$o$'' subscript in the following.

\subsection{Dirty limit and expansion of the quasiclassical Green's function}

Defining the direction of the Fermi momentum as $\vec e_F = \vec p_F/|\vec p_F|$, we expand the (linearized) quasiclassical Green's function to linear order in spherical harmonics,
\begin{equation}
    \ubar f \approx \ubar f_s + \vec e_F \cdot  \ubar{ \bm{f}}_p,
    \label{eq:diffusion}
\end{equation}
assuming that the strong impurity scattering causes the $s$-wave part to be much larger than the $p$-wave part, $\ubar f_s \gg \ubar{\bm{f}}_p$.
This is the so-called dirty limit, corresponding to quasiparticle diffusion.
In this expansion, the first term is even in momentum, whereas the second is odd.
We note that we can isolate the even part of the Green's function by averaging over the (solid angle of the) Fermi surface,
\begin{align}
    \int \frac{d\vec e_F}{2 \pi} \ubar f = \ubar f_s,
\end{align}
similarly, we can isolate the odd part by first multiplying with $\vec e_F$ and then performing the average
%averaging:
\begin{align}
    \int \frac{d\vec e_F}{2 \pi} \vec e_F \ubar f = \frac 12 \ubar{\bm f}_p .
\end{align}

In the following, we split \eq \eqref{eq:eilenberger_lin_ret_fo} into an even and an odd equation.
Performing the averages for the altermagnetic terms produces:
\begin{align}
    -\frac{\alpha}{2m} T_{ij}  
    \int \frac{d\vec e_F}{2 \pi}  p_{F, i} p_{F, j}
    \left[\ubar \tau_3, \ubar f_s + \vec e_F \cdot \ubar{\bm f}_p \right]_- 
    &= -\frac12 \frac{\alpha}{2m} p_F^2 \left[\ubar \tau_3, \ubar f_s \right]_- \text{Tr} T  = 0
    \label{eq:Eq42}
    \\
     \frac{\i \alpha}{2m}  \int \frac{d\vec e_F}{2 \pi}  T_{ij} 
    p_{F, i} \partial_{R_j} 
    \left[\ubar \tau_3, \ubar f_s + \vec e_F \cdot \ubar{\bm f}_p \right]_+ 
    &=
    \frac{\i \alpha v_F}{4} T_{ij} \partial_{R_j} 
    \left[ \ubar \tau_3,  \ubar f_{p, i} 
    \right]_+
    \\
    \frac{\alpha}{8m}  T_{ij} \partial_{R_i} \partial_{R_j} \int \frac{d \vec e_F}{2 \pi} \left[\ubar \tau_3, \ubar f_s + \vec e_F \cdot \ubar{\bm f}_p  \right]_-
    &= 
    \frac{\alpha}{8m}  T_{ij} \partial_{R_i} \partial_{R_j}  \left[\hat m, \ubar f_s \right]_-.
\end{align}
where we, in the final part of the first equation, used that the $T$-matrix from \eq \eqref{eq:T} is traceless and introduced the Fermi velocity $v_F = p_F / m$. 
Consequently, the even equation becomes
\begin{subequations}    \label{eq:even_eq_final}
\begin{align}
    &\frac{\i v_F}{2} \nabla_{\vec R}\cdot \vec{\ubar f}_p 
    +  2 E \ubar f_s
    + \frac{\i \alpha v_F}{4} T_{ij} \partial_{R_j} \left[ \ubar \tau_3,  \ubar f_{p, i} \right]_+ 
    + \frac{\alpha}{8m} T_{ij} \partial_{R_i} \partial_{R_j}  \left[\ubar \tau_3, \ubar f_s \right]_-
    = 0.
\end{align}
\end{subequations}
We now assume that impurity scattering is the dominating term such that the odd altermagnetic terms are negligible.
Using $ \check \Sigma_\text{imp} = - \i \check g_s / (2 \tau)$  for the impurity self-energy \cite{mortenSpinChargeTransport2003}, where $\tau$ is the elastic scattering time, we find
\begin{align}
    - [\hat \Sigma_\text{imp}, \hat g^R] &= 
    \frac{\i}{2 \tau} [\hat g_s^R, \hat g^R]_- = \frac{\i}{2 \tau} [\hat \rho_3 + \hat f_s, \hat \rho_3 + \hat f_s + \vec e_F \cdot \hat{\bm f}_p] \approx \frac{\i}{2 \tau} [\hat \rho_3, \vec e_F \cdot \hat{\bm f}_p]
    \\
    &=  \frac{\i}{2 \tau}\vec e_F \cdot [\hat \rho_3, \hat{\bm f}_p]_-
    = \frac{\i}{2 \tau}
    \vec e_F \cdot 
    \begin{pmatrix}
    \ubar 0 &  \ubar{\bm f}_p\\
    \ubar{\tilde{\bm f}}_p& \ubar 0
    \end{pmatrix}.
\end{align}
We multiply \eq \eqref{eq:eilenberger_lin_ret_fd} with $\vec e_F$ and average over the Fermi surface to find the odd equation:
\begin{align}
    \label{eq:gp}
    \bm f_p = - \tau v_F \nabla_{\vec R} \ubar f_s
\end{align}

Inserting \eq \eqref{eq:gp} into \eq \eqref{eq:even_eq_final} and multiplying with $\i$, we find the Usadel equation as two decoupled equations for the $2\times 2$ matrix $\ubar f$:
\begin{align}
\label{eq:lin_Usadel_final}
    D \nabla_{\vec R}^2  \ubar f = - 2 \i E  \ubar f - \frac{\alpha D}{2}  T_{ij} \partial_{R_i} \partial_{R_j} [\ubar \tau_3,  \ubar f]_+ 
    -  \frac{\i \alpha}{8m} T_{ij} \partial_{R_i} \partial_{R_j} [\ubar \tau_3,   \ubar f]_-,
\end{align}
where we defined the diffusion constant $D = \tau v_F^2 / 2 $.
The factor 2 in the denominator of the diffusion constant is because this is a 2D system. More generally, the diffusion constant reads $D = \tau v_F^2 / d$, where $d$ is the spatial dimension of the system.

\subsection{Singlet-triplet decomposition}
We follow Ref. \cite{Jacobsen2015} and introduce a singlet/triplet decomposition,
\begin{align}
    \ubar f = (f_s + \vec d \cdot \bm \sigma ) \i \ubar \sigma_2 = \begin{pmatrix}
        \i d_y - d_x & d_z + f_s\\
        d_z - f_s & \i d_y + d_x 
    \end{pmatrix},
\end{align}
where $f_s$ denotes the superconducting singlet amplitude and $\vec d$ the superconducting triplet amplitudes.
We calculate the second term on the right-hand side of \eq \eqref{eq:lin_Usadel_final} using this decomposition and find
\begin{align}
    \frac{\alpha D}{2} T_{ij} \partial_{R_i} \partial_{R_j} [\ubar \tau_3,  \ubar f]_+
    =  \alpha D T_{ij} \partial_{R_i} \partial_{R_j} 
    \begin{pmatrix}
        i d_y - dx  & 0\\
        0 & - i d_y - d_x
    \end{pmatrix}.
\end{align}
The third term takes the form
\begin{align}
    \frac{\i \alpha}{8m} T_{ij} \partial_{R_i} \partial_{R_j} [\ubar \tau_3, \ubar f]_-
    = \frac{\i \alpha}{4m} T_{ij} \partial_{R_i} \partial_{R_j} \begin{pmatrix}
        0 & d_z + f_s \\ f_s - d_z  & 0 
    \end{pmatrix} \, .
\end{align}
In total, the nontrivial (i.e. coupled) parts of \eq \eqref{eq:lin_Usadel_final} becomes 
\begin{subequations}
\begin{align}
    D \nabla^2 d_x = -2 \i E d_x  - \alpha D T_{ij} \partial_{R_i} \partial_{R_j} (i d_y) \\
    D \nabla^2 (i d_y) = -2 \i E (\i d_y)  - \alpha D T_{ij} \partial_{R_i} \partial_{R_j} d_x,
\end{align}
\label{eq:lin_Usadel_triptrip}
\end{subequations}
and 
\begin{subequations}
\label{eq:lin_Usadel_singtrip}
\begin{align}
\label{singlet_final}
    D \nabla^2 f_s = - 2 \i E f_s +  \frac{\i \alpha}{4m} T_{ij} \partial_{R_i} \partial_{R_j} d_z
    \\
    \label{triplet_final}
    D \nabla^2 d_z = - 2 \i E d_z +  \frac{\i \alpha}{4m} T_{ij} \partial_{R_i} \partial_{R_j} f_s.
\end{align}
\end{subequations}
It is convenient to introduce the quantities $f_{\uparrow} = \i d_y - d_x$ and $f_\downarrow = \i d_y + d_x$, and transform \eq \eqref{eq:lin_Usadel_triptrip} into:
\begin{subequations}
\label{eq:lin_Usadel_fupfdn}
\begin{align}
    D \nabla^2 f_\uparrow = -2 \i E f_\uparrow  +  \alpha D T_{ij} \partial_{R_i} \partial_{R_j} f_\uparrow\\
    D \nabla^2 f_\downarrow = - 2 \i E f_\downarrow  - \alpha D T_{ij} \partial_{R_i} \partial_{R_j}  f_\downarrow.
\end{align}
\end{subequations}
Here, $f_\uparrow$ and $f_\downarrow$ are condensate amplitudes for Cooper pair triplets with spin up and down, respectively. 
Similarly, it is convenient to introduce $f_1 = f_s + d_z$ and $f_2 = f_s - d_z$ and transform \eq \eqref{eq:lin_Usadel_singtrip}:
\begin{subequations}
\label{eq:lin_Usadel_f1f2}
\begin{align}
    D \nabla^2 f_1 = -2 \i E f_1  +  \frac{\i \alpha}{4m} T_{ij} \partial_{R_i} \partial_{R_j} f_1\\
    D \nabla^2 f_2= - 2 \i E f_2 - \frac{\i \alpha}{4m} T_{ij} \partial_{R_i} \partial_{R_j} f_2.
\end{align}
\end{subequations}
%These functions do not, however, have a natural interpretation, and one should think of this transformation only as a convenient way to decouple \eq \eqref{eq:lin_Usadel_singtrip}.
%After solving \eq \eqref{eq:lin_Usadel_f1f2}, one should transform back to the amplitudes $f_s$ and $d_z$ to gain a physical interpretation of the observable effects.

\section{Derivation of quasiclassical boundary conditions}

In this section, we derive boundary conditions for interfaces with altermagnets using a tunneling Hamiltonian. 
The derivation is done by adjusting the Kupriyanov-Lukichev boundary conditions, which can be done using the methodology of \cite{linder_prb_22}, with a current operator consistent with the altermagnet.

\subsection{The quasiclassical current operator in an altermagnet}

The effective altermagnetic Hamiltonian is
\begin{align}
    H = \frac{\vec p^2}{2m} + \frac{\alpha}{2m} \ubar \sigma_z  T_{ij} p_i p_j.
\end{align}
Hamilton's equation of motion implies that the velocity operator is spin- and momentum-dependent. It is on the form
\begin{equation}
    v_i = \frac{\partial H}{\partial  p_i} = \frac{ 1}{m} 
        \left(p_i + \alpha \ubar \sigma_z  T_{ij} p_j \right) .
\end{equation}
Here, we have used the chain rule and the symmetry of $T$ to write $\partial_{p_i} (T_{lj} p_l p_j ) = 2 T_{ij} p_j$.
For the two special cases $\theta = 0$ and $\theta = \pi /4$, the velocity operator reads
\begin{align}
\label{eq:velop}
    \vec v = 
    \begin{cases}
  \frac 1 m 
  \begin{pmatrix}
      ( 1 + \alpha \ubar \sigma_z) p_x  \\ ( 1 - \alpha \ubar \sigma_z) p_y
  \end{pmatrix},  & \theta = 0 \\
    \frac{1}{ m} 
    \begin{pmatrix}
        p_x +  \alpha \ubar \sigma_z p_y\\
        p_y +  \alpha \ubar \sigma_z p_x
    \end{pmatrix}, & \theta = \pi / 4.
\end{cases}
\end{align}
The quasiclassical current operator should be modified accordingly:
\begin{align}
    \check I & \propto
     \vec n \cdot \int dE \Tr \left\{ \hat \rho_3 \left( \check g_{L, s} 
    \vec v
    \check g_{L, s} \right) \right\}.
    \label{eq:I_AM1}
\end{align}
From \eq \eqref{eq:I_AM1} and \eq \eqref{eq:velop}, it is apparent that the boundary conditions are spin-dependent, and couple the $x$ and $y$ components in the case of $\theta = \pi / 4$.
The momentum components $p_x$ and $p_y$ are more generally coupled for any $\theta \neq 0$.

\subsection{Modifying the Kupriyanov-Lukichev boundary conditions}

The spin-active Kupriyanov-Lukichev boundary conditions in quasiclassical theory are expressed as 
\cite{kupriyanov_zetf_88, ouassou_scirep_17}
\begin{align}
\label{eq:bdc_final_L2}
    2  \check I_R
    =  G_0 [  \check g_{L}, \check g_{R}]_- + \i G_\phi [ \check g_{R}, \hat m_x]_-,
\end{align}
where the matrix current can be expressed in terms of the right-hand side material
\begin{align}
\label{eq:matrix_curr}
    \check I_R 
    % =  G_L  L_L  \check g_{L} ( \vec n \cdot \nabla_{\vec R} )\check g_{L}  
    = G_R L_R  \check g_{R} ( \vec n \cdot \nabla_{\vec R} )\check g_{R} ,
\end{align}
where $\check g_{L (R)}$ is the quasiclassical Green's function to the left (right), and we have included a possible spin polarization in the $x$-direction.
In Eq. \eqref{eq:matrix_curr}, $G_L$ is the bulk conductance of material $L$, and $L_L$ is the length of the material.
For the interface where the AM is on the left side, the boundary condition is obtained by letting $\check{I}_R \to  -\check{I}_L$ on the left-hand side and $\check{g}_L \leftrightarrow \check{g}_R$ on the right-hand side of the boundary condition equation.
In the following, we denote the conductance of the altermagnet as $G$ and its length as $L$.
$G_0$ denotes the interface conductance, and $G_\phi$ is the (spin-mixing) conductance of the magnetic term. We have assumed the interface polarization to be in the $x$-direction.
% We derive the boundary conditions for a system with an altermagnet to the right and a superconductor to the left, but the reverse case is easily obtained by $n_i \rightarrow - n_i$ and $L \leftrightarrow R$.
The Kupriyanov-Lukichev boundary conditions are derived using the quasiclassical current operator and are modified under the presence of altermagnetism in, e.g., the right region:
\begin{equation}
     2  G L  \check g_{R}   ( \nabla_{ R_i} + \alpha \ubar \sigma_z T_{ij} \partial_{R_j}) \check g_{R}
     =G_0 [  \check g_{L}, \check g_{R}]_- + \i G_\phi [ \check g_{R}, \hat m_x]_-  ,
\end{equation}
where the repeated Latin indices imply the sum over the Cartesian components.
The corresponding equation for the Green's function on the right side of the interface follows from a substitution $\vec n \rightarrow - \vec n$ and $\check g_{L} \leftrightarrow \check g_{R}$, as well as a modification of the current operator in the altermagnet due to the velocity operator being different from a normal metal.

\subsection{Linearization}
We linearize the (retarded component of the) Kupriyanov-Lukichev boundary conditions in \eq \eqref{eq:bdc_final_L2}, i.e. $\hat g_R = \hat \rho_3 + \hat f_R$, and include terms $\mathcal O (\hat f_R )$ on the left-hand side, and terms $\mathcal O (\hat f^0_R)$  on the right-hand side:
\begin{align}
\label{eq:bdc_lin1}
    2 G L  \hat \rho_3 \left(\partial_{R_i} + \alpha \ubar \sigma_z T_{ij} \partial_{R_j} \right) \hat f_R = G_0 [\hat g_L, \hat \rho_3] + \i G_\phi [ \hat f_{R}, \hat m_x],
\end{align}
where $\hat f_{L(R)}$ has the symmetries described in \eq \eqref{eq:f_hat}.
To see how the boundary condition affects the diagonal part $\ubar f_d$ and the off-diagonal part $\ubar f_o$, we multiply \eq \eqref{eq:bdc_lin1} with $\hat \rho_3$ from the left and find:
\begin{align}
     2 G L   n_i  \left(\partial_i + \alpha \ubar \sigma_z T_{ij} \partial_j \right)
    \begin{pmatrix}
        \ubar f_d & \ubar f_o \\[4pt]
        -\ubar{\tilde f_o} & -\ubar{\tilde f_d}
    \end{pmatrix}_R
    = -2 G_0 
    \begin{pmatrix}
        0 & \ubar f_o \\[4pt]
        -\ubar{\tilde f_o} & 0
    \end{pmatrix}_L
    + \i G_\phi 
    \begin{pmatrix}
        [\ubar f_d, \sigma_x]_- & [\ubar f_o, \sigma_x]_-\\[4pt]
        [\ubar{\tilde f_o}, \sigma_x]_- & [\ubar f_d, \sigma_x]_-
    \end{pmatrix}_R .
\end{align}
The boundary conditions for $\ubar f_d$ and $\ubar f_o$ are decoupled to the first order in $\hat f$. We will not need to consider the diagonal parts in more detail, as described in Sec.\ \ref{section:norm_metal}, and drop the subscript in $\ubar f_o$ from here.

The linearized Kuprianov-Lukichev boundary condition for the anomalous Green's function for any altermagnetic orientation reads
\begin{align}
\label{bdc4}
     2 G L  \left(\partial_{R_i} + \alpha \ubar \sigma_z T_{ij} \partial_{R_j} \right)
    \ubar f_R
    =- 2 G_0 
    \ubar f_L
    + \i G_\phi [ \ubar f_{R}, \ubar \sigma_x]_-.
\end{align}

It is instructive to consider some special cases of this. We consider an interface along the $x$-direction and the $\theta = 0$ and $\theta = \pi / 4$ orientations of the altermagnet separately. 
We assume an $s$-wave singlet BCS superconductor in $L$, i.e. 
\begin{equation}
    \ubar f_L = \begin{pmatrix}
        0 & f_{\mathrm{BCS}}\\
        - f_{\mathrm{BCS}} & 0
    \end{pmatrix},
\end{equation}
where $f_{\mathrm{BCS}} $ is the (bulk) value of the anomalous Green's function in the superconductor. Moreover, the spin-active term becomes
\begin{align}
    [ \ubar f_{R}, \ubar \sigma_x]_- =
    2 \begin{pmatrix}
        f_s & - d_x\\
        d_x & -f_s
    \end{pmatrix}_R.
\end{align}
We detail the two special cases for the altermagnetic orientations in the following.

\subsubsection{\texorpdfstring{$\theta = 0$}{TEXT} orientation}
Employing the same singlet-triplet decomposition as in Sec.\ \ref{section:Usadel_linearization}, we find that the boundary condition in the $\theta = 0$ orientation becomes
\begin{subequations}
\begin{align}
     L ( 1 + \alpha) \partial_{R_x} f_{\uparrow  , R} = \i \frac{G_\phi}{G} f_{s, R}\\
     L ( 1 - \alpha) \partial_{R_x} f_{\downarrow, R} =  -\i \frac{G_\phi}{G}  f_{s, R},
\end{align}
\end{subequations}
for the equal-spin Cooper pairs, and 
\begin{subequations}
\begin{align}
    L ( 1 + \alpha) \partial_{R_x} f_{1, R} = -\frac{G_0}{G} f_{\mathrm{BCS}} - \frac{G_\phi}{G} 2 d_{x, R} \\
    L ( 1 - \alpha) \partial_{R_x} f_{2, R} = -\frac{G_0}{G} f_{\mathrm{BCS}} + \frac{G_\phi}{G} 2 d_{x, R},
\end{align}
\end{subequations}
for the singlet/triplet amplitudes $f_{1/2} =  f_s \pm d_z$.
These equations show that the singlet and triplet amplitudes mix at the boundary and that the mixing scales with the altermagnetic strength $\alpha$.

\subsubsection{\texorpdfstring{$\theta = \pi / 4$}{TEXT} orientation}

In the $\theta=\pi/4$ orientation, the Kupriyanov-Lukichev boundary conditions include gradients along both $x$ and $y$:
\begin{subequations}
\begin{align}
    L (\partial_{R_x} + \alpha \partial_{R_y})  f_{\uparrow , L} =  \i \frac{G_\phi}{G} f_{s, R} \\
    L (\partial_{R_x} - \alpha \partial_{R_y}) f_{\downarrow, L} =   -\i \frac{G_\phi}{G} f_{s, R},
\end{align}
\end{subequations}
for the equal-spin Cooper pairs, and 
\begin{align}
    L (\partial_{R_x} + \alpha \partial_{R_y}) f_{1, L} = -\frac{G_0}{G} f_{\mathrm{BCS}} - \frac{G_\phi}{G} 2 d_{x, R}\\
    L (\partial_{R_x} - \alpha \partial_{R_y}) f_{2, L} = -\frac{G_0}{G} f_{\mathrm{BCS}} + \frac{G_\phi}{G} 2 d_{x, R}
\end{align}
for the singlet/triplet amplitudes $f_{1/2} =  f_s \pm d_z$. 

\section{Solution strategy}

This section outlines and summarizes the numerical solution strategy used to solve the derived Usadel equations with the corresponding Kupriyanov-Lukichev boundary conditions.

We solve for the singlet/triplet amplitudes in the normal metals connected to the altermagnet, as described in Sec.\ \ref{section:norm_metal}, and use these to obtain observable quantities such as magnetization. 
To do this, we solve the Usadel equations \eqref{eq:lin_Usadel_fupfdn} and \eqref{eq:lin_Usadel_f1f2} together with the boundary conditions \eqref{bdc4} to determine the amplitude in the altermagnet.
The four different Cooper pair anomalous Green's function amplitudes are coupled. However, the equal-spin triplet amplitude $d_x$ is negligible in the weak proximity regime. We solve for $f_{1, 2}$ and use the result in the boundary conditions for the equal-spin triplets. n this limit, the equations for $f_{1, 2}$ and $f_{\uparrow, \downarrow}$ decouple. In this paper, we solve these equations using a simple finite difference scheme on a square geometry. 

Finally, we use the amplitudes in the altermagnet to find the amplitudes in the normal metal using Eq. \eqref{eq:C}. This allows us to compute the magnetization in the normal metal.

\bibliography{masterref}